\documentstyle[twoside,fleqn,espcrc2,psfig]{article}

\newcommand{\AmS}{{\protect\the\textfont2
  A\kern-.1667em\lower.5ex\hbox{M}\kern-.125emS}}
\newcommand{\be}{\begin{equation}}
\newcommand{\ee}{\end{equation}}
\def\reff#1{(\ref{#1})}

\newcommand{\ba}{\begin{eqnarray}}
\newcommand{\ea}{\end{eqnarray}}
\def\spose#1{\hbox to 0pt{#1\hss}}
\def\ltapprox{\mathrel{\spose{\lower 3pt\hbox{$\mathchar"218$}}
 \raise 2.0pt\hbox{$\mathchar"13C$}}}
\def\gtapprox{\mathrel{\spose{\lower 3pt\hbox{$\mathchar"218$}}
 \raise 2.0pt\hbox{$\mathchar"13E$}}}

\title{Color-Coulomb Force Calculated from Lattice Coulomb Hamiltonian}

\author{Attilio Cucchieri\address{Department of Physics, New York University,
             4 Washington Place, New York, NY 10003, USA}\thanks{Poster
           presented by A.~Cucchieri.} 
        \addtocounter{address}{-1} $\!\!\!$
        and 
        Daniel Zwanziger\addressmark}

\begin{document}

\begin{abstract}
     The static color-Coulomb potential is calculated as the
solution of a non-linear integral equation. This equation
has been derived recently as a self-consistency condition
which arises in the Coulomb Hamiltonian formulation of
lattice gauge theory when the restriction to the interior of
the Gribov horizon is implemented.
The potential obtained  is
in qualitative agreement with expectations, being
Coulombic with logarithmic corrections at short range and
confining at long range. The values obtained for the string
tension and $\Lambda_{\overline{MS}}$
are in semi-quantitative agreement
with lattice Monte Carlo and phenomenological
determinations.
\end{abstract}

\maketitle

\section{INTRODUCTION}

The lattice Coulomb-gauge Hamiltonian has recently been
derived from the transfer matrix of Wilson's Euclidean
lattice gauge theory \cite{ZCoul}.
The physical configuration space (no Gribov copies)
is restricted to the fundamental modular region $\Lambda$ 
of the minimal Coulomb gauge ---
{\em i.e.}\ to
the set of absolute minima of a Morse function
on the gauge orbits.
This restriction is implemented
by the effective Hamiltonian
\be
{\cal H}_{eff}\,\equiv\,{\cal H}_{coul}\,+\,\gamma_{0}\,G
\;\mbox{,}
\label{eq:Heff}
\ee
where ${\cal H}_{coul}$ is
a lattice
analog of
the Christ-Lee Hamiltonian \cite{Lee}, and
$G$ is
the (3-dimensional) {\em horizon function} \cite{ZCoul}. Here $\gamma_{0}$ is a
new thermodynamic parameter whose value is determined by
the {\em horizon condition}
$\langle\,G\,\rangle / V \,=\,0$, where the expectation-value is
calculated in the ground state of ${\cal H}_{eff}$.
The horizon condition contradicts the
usual perturbative expansion, but it is consistent with an
expansion in powers of $g_{0}$ and the Ansatz \cite{ZCoul}
\be
\left( {\cal M}^{-1} \right)^{a b}(x, y; A) \, = \,
    g_{0}^{- 1}\, \delta^{a, b}\, u(x - y) \,+\,\ldots
\label{eq:Mofg0}
\;\mbox{,}
\ee
where ${\cal M} \equiv - D \cdot \nabla$ is the
3-dimensional Faddeev-Popov operator that appears
in ${\cal H}_{coul}$ \cite{Lee}. To lowest order in
$g_{0}$, the Schwinger-Dyson equation for the
Faddeev-Popov propagator
becomes
an equation for the function $u$ which, in the continuum limit,
reads in momentum space \cite{ZCoul}
\ba
\frac{1}{u(q)} \!\! &=& \!\! \frac{n}{( 2 \pi )^{3}}
   \int d^{3}k\, D^{0}(k)\biggl\{ \left[\, q^{2} -
        ( q \cdot k)^{2} k^{- 2}\,\right] \nonumber \\
& & \quad \quad \times \left[\, u(k)\,-\,u(k + q)\,\right] \biggl\}
\;\mbox{,}
\label{eq:inteq}
\ea
where $D^{0}(k)$ is the equal-time gluon propagator, to
zeroth order in $g_{0}$.

The kernel $D^{0}(k)$ is an expectation-value
calculated in the ground state
$\Psi_{0}[u]$ of
the effective Hamiltonian ${\cal H}_{eff}$
to zeroth order in $g_{0}$. With neglect of
dynamical fermions, two
terms
contribute to
\be
{\cal H}^{0}_{eff} \,=\, {\cal H}^{0}_{ho}\,+\, {\cal H}^{0}_{ci}
\;\mbox{,}
\label{eq:Heff0}
\ee
a harmonic-oscillator-like term
and a Coulomb-interaction one,
\be
{\cal H}^{0}_{ci} \equiv [ \,2\,( 2 \pi )^{3}\,]^{- 1}\,
  \int d^{3}k \, {\tilde \rho}^{a}(- k)\,v(k)\,
    {\tilde \rho}^{a}(k) \; \mbox{.}
\label{eq:Hci}
\ee
Here $v(k) \equiv k^{2} u^{2}(k)$, and
${\tilde \rho}^{a}(k)$ is the Coulomb-gauge color-charge
density defined in \cite{ZCoul,Lee}.
We cannot hope
to solve this problem exactly because it involves the
color-Coulomb interaction of dynamical gluons.
We use instead the ground state $\Psi_{0}^{(0)}[u]$
of the harmonic oscillator Hamiltonian only and we obtain
\cite{ZCoul}
\be
D^{0 (0)}(k) \,=\,\frac{1}{ 2\,\omega(k)}
   \,=\,\frac{1}{ 2\,\sqrt{\,k^{2}\,+\,
  \mu^{4}\,u(k)\,}}
\label{eq:D00}
\;\mbox{.}
\ee
An estimate of the leading correction gives a renormalization
of the length scale, but does not otherwise change
$D^{0}(k)$ qualitatively.

Let $w(k) \equiv u^{(0)}(k)$ be the solution to \reff{eq:inteq}
with this kernel.
If we define [in the $SU(n)$ case]
\be
w(q\mbox{,}\,\mu) \equiv \frac{ {\hat g}(q\mbox{,}\,\mu)}{q^{2}}
\equiv \frac{1}{q^{2}\, n^{1 / 2}}
g\left(\frac{q\,n^{- 1 / 8}}{ \mu}\right)
\;\mbox{,}
\label{eq:grescale}
\label{eq:defofw}
\ee
then \reff{eq:inteq} becomes
\ba
g^{- 1}(q) \!\! &=& \!\! \frac{1}{8 \pi^{2}}
      \int_{0}^{\infty} dk\,\int_{-1}^{1} dz\,\Biggl\{ \nonumber \\
& & \,\,\,
\frac{ k^{3}\left(\,1 - z^{2}\,\right)}{
\sqrt{k^{4} + g(k)}}\,
\left[\frac{g(k)}{k^{2}}-\frac{g(p)}{p^{2}}\right]\Biggl\}
\label{eq:int.eqforg}
\ea
with $p^{2} \,\equiv\, q^{2} \,+\,k^{2}\,+\,2\,k\,q\,z$.
It is easy to check \cite{ZCoul} that
\ba
g(q)&=&B\,q^{- 4 / 3}
\label{eq:gto0} \\
B&=&\left[\,\frac{\Gamma( 16 / 3 )\,\pi^{2}}{\Gamma( 8 / 3 )
\, \Gamma(2 / 3 )}\,\right]^{2/3}
\label{eq:Bvalue}
\ea
is a self-consistent solution of the integral equation
\reff{eq:int.eqforg} in the IR limit.
In \cite{ZCoul,CZ} it has been proven that,
as $q \to \infty$, the function ${\hat g}(q)$, defined in \reff{eq:grescale},
is given by
\ba
{\hat g}^{- 2}(q) \!\! &=& \!\!  n \, ( 6\,\pi^{2} )^{- 1}
            \,\left[\,\log{( q^{2} m_{1}^{- 2})}
   \right. \nonumber \\
& & \left.  \, \, \, +\,
 3^{- 1}\,\log{\log{( q^{2} m_{2}^{- 2} )}}\,\right]
\,+\,\ldots
\label{eq:gasalpha1}
\;\mbox{,}
\ea
where $m_{1}$ and $m_{2}$ are unknown constants.

The function ${\hat g}^{2}(q)$ plays the role
of a running coupling constant $4\,\pi\,\alpha(q)$
and has the UV behavior predicted 
by the perturbative renormalization group, although
the coefficient $b_{0} \,=\, n / (6\,\pi^{2}) $ does not
have the expected value $(11\,n) / (48\,\pi^{2})$.
As explained in \cite{ZCoul}, one recovers the
correct value for this coefficient from
terms in ${\cal H}_{eff}$
of higher order in $g_{0}$.	

\section{A TRIAL SOLUTION FOR THE INTEGRAL EQUATION}
\label{Sec:trial}

In order to find an approximate numerical solution for the
integral equation \reff{eq:int.eqforg} we
follow the approach in \cite{BBZ} and simplify the problem
by using a trial solution depending on a set of parameters
\cite{Athesis}.
We have adopted the form
\ba
g(q) & = &\frac{B\,q^{- 4 / 3}}{
     1\,+\,\left(
          q / q_{0} \right)^{( 2 + \nu)}}
           + \frac{\pi \sqrt{6}}{1\,+\,\left(
            q_{0} / q \right)^{( 2 + \nu)}}
     \, \biggl\{ \nonumber \\
& &\,\,\, \, \quad
\log{(1 + q^{2} m_{1}^{- 2} )} \,+\,
3^{- 1}\,\log \biggl[ \, 1 \nonumber  \\
& & \quad \qquad +\,
 \log{(1\,+\, q^{2}  m_{2}^{- 2} )}
\,\biggl]\,\biggl\}^{- 1 / 2}
\label{eq:gtrial}
\ea
which has the asymptotic behaviors \reff{eq:gto0} and \reff{eq:gasalpha1}.
This function is positive,
monotonically decreasing
and depends on the four parameters $m_{1}$,
$m_{2}$, $q_{0}$ and $\nu$.

We tuned these parameters
by evaluating \reff{eq:int.eqforg} numerically using
the set $I$ of values of $q$ described below.
To compare the results obtained for
different sets of parameter values we
used the quantity
\be
M(I)  \equiv \,\max_{q \in I} \, \Delta g(q) \,
  \equiv \, \max_{q \in I} \, \left|\,1 - \frac{g_{out}(q)}{
                       g_{in}(q)} \,\right|
\;\mbox{.}
\ee
Here $g_{in}(q)$ is used on the r.h.s. of \reff{eq:int.eqforg}
and $g_{out}(q)$ is the result obtained on the l.h.s.
The set $I$ consisted of $105$ points
in the interval $[\,10^{- 7}\mbox{,}\,10^{6}\,]$.
The accuracy of the numerical
integration was fixed to five parts in $10^{3}$.
The best values we found,
$q_{0} = 1.87$, $m_{1} = 1.54$, $m_{2} = 2.74$ and
$\nu = -0.06$,
give $ M(I) \, =\,\,\,0.01826$. 
We also evaluated $ A(I) \equiv \langle\,\Delta g(q) \,\rangle$,
where the average is taken over all the points in $I$.
We obtained $A(I) = 0.00752 \,\pm\, 0.00054$.

In order to test our solution we checked \cite{Athesis}:
$(1)$ that the theoretical
behavior at small $q$ is satisfied;
$(2)$ that our result is independent of the
set of points $I$ used for tuning the four parameters; and
$(3)$ that our solution is stable,
namely if we use different values for the parameters, the
corresponding output points $g_{out}(q)$ move in the
direction of our solution.

\section{THE POTENTIAL AND THE FORCE}
\label{Sec:potandfor}

The potential $v(k)$ which appears in \reff{eq:Hci}
has the Fourier transform
(apart from an additive constant)
\be
V(r) = 
     -\frac{n^{2} - 1}{2\,n^{2}} \frac{1}{2\,\pi^{2}}
                           \int_{0}^{\infty}
  dk\, g^{2}(k) \frac{\sin{(\,k\,r\,)}}{k\,r}
\;\mbox{.}
\label{eq:Vr3}
\ee
This gives \cite{ZCoul} a force $f(r) \equiv - \,\partial_{r}\, V(r)$
that at large separations
goes as
\be
f(r)\,\approx\,-\, \frac{n^{2} - 1}{2\,n^{2}}\, \frac{1}{4\,\pi}
           \,\frac{10\, B^{2}\, r^{2/ 3}}{3\,\Gamma( 11 / 3 )}
\;\mbox{,}
\label{eq:flarger1}
\ee
and a potential energy that grows as $r^{5 /3}$.

From \reff{eq:gasalpha1} we obtain
the limiting behavior of the force at small $r$,
\ba
f(r)&\approx& 3\,\pi\,( n^{2} - 1 ) \,( 4\,n^{2}\,r^{2} )^{- 1}
  \,\biggl[ \,
   \log{( \Lambda_{R}^{2}\,r^{2} )} \nonumber \\
& & \qquad \,\, \,\, +\, 3^{- 1}
   \,\log{\log{( \lambda^{2}\, r^{2})}}\,\biggl]^{- 1}
\;\mbox{,}
\label{eq:fsmallr}
\ea
where $\Lambda_{R} \equiv  e^{\gamma \,-\,1} \, m_{1}$,
$\gamma$ is the Euler constant,
and $\lambda$ is a constant.

\section{RESULTS AND CONCLUSIONS}

The potential $V(r)$ which we have calculated appears in the quantum
field theoretic Hamiltonian ${\cal H}_{eff}$.
A related quantity is the ground state
energy $E(r)$ of ${\cal H}_{eff}$
in the presence of a pair of external quarks at separation
$r$, ${\cal H}_{eff} \Psi_{0}
 = E(r) \Psi_{0}$, which is gauge-invariant.  They differ by the QCD
analog of vacuum polarization. However because of asymptotic freedom we
expect that $V(r)$ behaves like $E(r)$
for small $r$, once a physical length scale is
adopted. We also
expect that $V(r)$ and $E(r)$ differ significantly at large $r$.
The quantity $E(r)$ is known phenomenologically in the range
$0.2\, \mbox{fm} \ltapprox r \ltapprox 0.8\, \mbox{fm}$.
It is believed, consistent with lattice gauge theory
calculations \cite{Sch}, that in the absence of dynamical gluons
$E(r) = K r$,
holds asymptotically at large $r$, where $K$ is the string tension. The
asymptotic behavior $V(r) \sim r^{5/3}$, is the color-electric field energy of
two
superposed spherically symmetric color-electric fields of a quark and
anti-quark at separation $r$.
Although the power $5 / 3$ may be an artifact
of the approximation, it is to be expected that, for any power
that exceeds unity,
the ground
state wave function $\Psi_{0}( A^{tr} )$ adjusts itself so that at large $r$ the
color-electric field is contained in a flux tube, which gives a lower
energy, that rises linearly with $r$	.

As a test of these ideas and of the approximations made, we shall
directly compare our results for $V(r)$ with phenomenological fits
to $E(r)$, to see if there is a range of ``small'' $r$ for which $V(r)$
behaves like $E(r)$. To this end,
we consider
two phenomenological models:
the Cornell potential \cite{Eich} and the
Richardson potential \cite{Rich},
which give a good fit to
the $c {\overline c}$ and $b {\overline b}$
spectra.

From the analytic expression \reff{eq:gtrial},
we evaluate (numerically) $\,V(r)$ and $f(r)$.
To make a connection between these dimensionless
quantities and the real world, we fix the length scale
by using Sommer's dimensionless phenomenological relation \cite{Sommer}
\be
r^{2}_{0}\,f(r_{0}) = - 1.65
\label{eq:r0}
\;\mbox{,}
\ee
which holds for the Cornell force at
$r_{0}\,a \equiv R_{0}\,=\,0.49\,\mbox{fm}
\,=\, 2.48 \,(\mbox{GeV})^{- 1}$.

From the value of $a$ in $(\mbox{GeV})^{- 1}$ we obtain
the string tension \cite{Athesis}
\be
\sqrt{\,\sigma\,} \equiv \sqrt{\min_{r}\left[\,- f(r) \,\right]}\,a^{- 1}
\,=\,518\,\mbox{MeV}
\label{eq:stringten}
\;\mbox{.}
\ee
[Because $- f(r)$ increases like
$r^{2 / 3}$ when $r$ goes to infinity,
we cannot use the standard definition $\sigma
\equiv  \lim_{r \to \infty}\,- f(r)$.
Equation \reff{eq:stringten} defines
$\sigma$ where $f^{'}(r) = 0$, {\em i.e}\ where the
potential is approximately linear.]
It is not easy to estimate an uncertainty for the string tension.
However, its value seems to depend very weakly on the values
of the parameters of our trial solution \cite{Athesis}.

If
we identify the parameter $\Lambda_{R}$ in
\reff{eq:fsmallr} with the corresponding physical parameter \cite{Billoire},
then from
\be
\Lambda_{\overline{MS}} \,=\, \Lambda_{R} \, e^{- \gamma + 1 - 31 / 66}
\label{eq:LambdaMS}
\ee
we obtain $\Lambda_{\overline{MS}} =
124\,\pm\,12\,\mbox{MeV}$ (see \cite{Athesis}).

In Figure \ref{FIG:plot_4forze}a we plot our result for
$f(r)$, the Cornell forces and the Richardson one.
Our force gets its maximum value for a separation
of about $2.5 (\mbox{GeV})^{- 1}$ and it is almost constant
up to $4 (\mbox{GeV})^{- 1}
\approx 0.8\,\mbox{fm}$, the variation being of order
$12 \%$.

If, instead of \reff{eq:r0}, we use
\be
r^{2}_{0}\,f(r_{0}) \,=\, - 1.35
\label{eq:r0bis}
\;\mbox{,}
\ee
which holds for the Richardson force at
$R_{0} = 2.48 (\mbox{GeV})^{- 1}$,
we obtain
$\sqrt{\,\sigma\,} = 468\,\mbox{MeV}$,
$\Lambda_{\overline{MS}} = 118\,\pm\,12\,\mbox{MeV}$
and the plot shown in Figure \ref{FIG:plot_4forze}b.
In this case the agreement
is even better:
in fact, our force reaches its maximum value for a separation
of about $2.75 (\mbox{GeV})^{- 1}$, and its variation
at $4 (\mbox{GeV})^{- 1}$ is of order $8 \%$.

The fit is surprisingly good.
The force is in qualitative agreement
with phenomenological models, and
the values obtained for the string
tension and $\Lambda_{\overline{MS}}$ are in semi-quantitative agreement
with lattice Monte Carlo and phenomenological
determinations (see \cite{Sch,Eich} and \cite{Particle}).

It would appear that, with Sommer's normalization
\reff{eq:r0} or \reff{eq:r0bis},
the approximate equality
$V(r) \approx E(r)$ extends to the range $r \ltapprox 0.8 \,\mbox{fm}$,
and moreover that the approximations made in our calculation of
$V(r)$ do not qualitatively destroy this agreement. Although there is no
{\em a priori}
reason to expect that vacuum polarization of gluons should not be
important in this range, this may not be so surprising after all.
For although
vacuum polarization of quarks does
``break'' the string,
this is not yet manifest for $r \ltapprox 0.8 \, \mbox{fm}$.

\begin{figure}[t]
\psfig{figure=plot_4forze.ps,height=2.8in}
\vspace{0cm}
\psfig{figure=plot_4forze.bis.ps,height=2.8in}
\vspace{-1cm}
\caption{~Plot of: (i)
         our force $f(R)$ in the
         $SU(3)$ case (the curve which is decreasing at large
         $R$), (ii)
         the forces (for the $c {\overline c}$ and the
         $b {\overline b}$ cases)
         derived from the Cornell potential (the two curves
         very close to each other), and (iii)
         the Richardson
         force (the highest curve).
         In the first case we set the length scale
         by using Sommer's proposal \protect\reff{eq:r0}, while in the  second
         case we use the
         condition \protect\reff{eq:r0bis}.}
\label{FIG:plot_4forze}
\end{figure}


\begin{thebibliography}{9}

\bibitem{ZCoul} D.\ Zwanziger, Lattice Coulomb Hamiltonian
                and Static Color-Coulomb Field,
                hep-th/{\bf 9603203}.

\bibitem{Lee} N.\ Christ and T.\ D.\ Lee, Phys.\ Rev.\ {\bf D22}
              (1980) 939. 

\bibitem{CZ} A.\ Cucchieri and D.\ Zwanziger, Static
             Color-Coulomb Force, hep-th/{\bf 9607224},
             NYU-ThPhCZ7-22-96 preprint.

\bibitem{BBZ} M.\ Baker, J.\ S.\ Ball and F.\ Zachariasen,
              Nucl.\ Phys.\ {\bf B186} (1981) 560.

\bibitem{Athesis} A.\ Cucchieri, Numerical results in minimal lattice
      Coulomb and Landau gauges: color-coulomb potential and gluon and ghost
      propagators, PhD thesis, New York University (May 1996).

\bibitem{Sch} G.\ S.\ Bali and K.\ Schilling, Nucl.\ Phys.\  {\bf B} (Proc. Suppl.)
              {\bf 34} (1994) 147.

\bibitem{Eich} E.\ Eichten et al., 
               Phys.\ Rev.\ Lett. {\bf 34}
               (1975) 369;
               E.\ Eichten et al.,
               Phys.\ Rev.\  {\bf D21}
               (1980) 203;
               E.\ Eichten and F.\ Feinberg, Phys.\ Rev.\  {\bf D23}
               (1981) 2724.

\bibitem{Rich} J.\ L.\ Richardson, Phys.\ Lett.\  {\bf B82} (1979) 272.

\bibitem{Sommer} R.\ Sommer, Nucl.\ Phys.\  {\bf B411} (1994) 839.

\bibitem{Billoire} A.\ Billoire, Phys.\ Lett.\  {\bf B104} (1981) 472.

\bibitem{Particle} Particle Data Group,
                   Review of Particle Properties,
                   Phys.\ Rev.\  {\bf D50} (1994) 1173.

\end{thebibliography}
\end{document}